\documentclass[twocolumn,amsmath,amssymb,showpacs,prb]{revtex4}
\usepackage{graphicx}
\usepackage{epsf}
\newcommand{\be}{\begin{equation}}
\newcommand{\ee}{\end{equation}}
\newcommand{\bea}{\begin{eqnarray}}
\newcommand{\eea}{\end{eqnarray}}

\newcommand{\p}{\partial}
\newcommand{\s}{\sigma}

\newcommand{\la}{\langle}
\newcommand{\ra}{\rangle}
\newcommand{\rd}{\mbox{d}}
\newcommand{\ri}{\mbox{i}}
\newcommand{\re}{\mbox{e}}

\def\nn{\nonumber\\}

\def\up{\uparrow}
\def\da{\downarrow}
\def\r#1{(\ref{#1})}
\def\tp{\tilde{t}_\perp}
\def\tr{\rm Tr}
\def\z0{{\cal Z}}
\def\tm{\bar{t}_0}
\def\dt{\delta t}
\def\bl{{\bf l}}
\def\bm{{\bf m}}
\def\bn{{\bf n}}

\def\zh{\widehat{Z}}
\begin{document}

\title{A strange metal with a small Fermi surface and strong collective excitations}
\author{F.H.L. Essler$^1$ and A.M. Tsvelik$^2$}
\affiliation{$^1$ Department of Physics, University of Oxford, 1 Keble
Road, Oxford OX1 3NP, UK;\\  
$^2$ Department of Physics, Brookhaven National Laboratory,
Upton, NY 11973-5000, USA}

\begin{abstract}
We develop a theory of a hybrid
state, where quasi-particles coexist with strong collective modes, taking as a 
starting point  a model of infinitely many one-dimensional Mott insulators 
weakly coupled by interchain tunneling. This
state exists at an intermediate temperature range and undergoes an
antiferromagnetic phase transition at temperatures much smaller than
the Mott-Hubbard gap. The most peculiar feature of the hybrid state is
that its Fermi surface volume is unrelated to the electron density.   
We present a self-consistent derivation of the low energy effective
action for our model.

\end{abstract}


\maketitle

\section{Introduction}
 
The presence of strong collective modes interacting with quasi-particles
is a distinctive feature of many strongly interacting systems such as
'bad' metals, weakly doped Mott insulators (such as the cuprates) and
heavy fermion materials. This interaction is believed to result in a
variety of unusual phenomena observed in these systems such as the
violation of the Mott-Regel limit in the temperature dependence of the
electrical resistivity of bad metals or the absence of quasi-particle
peaks in the spectral function of the cuprates. The lack of
non-perturbative techniques in dimensions higher than one makes a
detailed theoretical description of these phenomena quite challenging.  
One sucessful approach has been developed by Chubukov, Schmalian and Abanov,
who have studied the so-called spin-fermion model put forward by
D. Pines \cite{pines}. This model is semi-phenomenological and
postulates the existence of a strong, coherent, collective mode, which
interacts with quasi-particles located in the vicinity of a Fermi
surface. This model is reviewed comprehensively in
Refs [\onlinecite{chub2}] and has proven quite successful in explaining 
various properties of the cuprates. However, a derivation from a
microscopic Hamiltonian is lacking. 

In this paper we provide a microscopic derivation of a model in the
same class as the spin-fermion model of Pines and Chubukov. Namely, we
continue to develop a theory of a hybrid state combining features of a
Landau Fermi liquid and a Mott insulator suggested in
Ref.[\onlinecite{gf2}]. This state is characterized by the coexistence
of well-developed collective modes with quasi-particles. The latter
ones  have a small Fermi surface (SFS), the volume of which is
unrelated to the total number of electrons.   
By definition, the Fermi surface (FS) is small if its volume is less
than the maximum volume allowed by Luttinger's theorem 
\cite{Lutt,AGD,Dz,AltChub}. The existence of such a state does not
contradict Luttinger's theorem since the latter, contrary to popular
belief, does not fix the volume of FS. Instead the theorem states that
the electron density $n$ is proportional to the volume of phase space
enclosed by the surface where the single electron Green's function
changes its sign
\bea
n = \frac{2}{(2\pi)^d}\int_{G(\omega = 0,{\bf k}) > 0}\rd^d k \ .
\label{Lut}
\eea 
When the Green's function has zeroes, the Fermi surface constitutes
only a part of this surface, namely the one where $G(0,{\bf k})
\rightarrow \infty$. Hence Luttinger's theorem (\ref{Lut}) does
not even require the existence of a Fermi surface: the Green's
function may only have zeroes and no poles, as it is the case 
for  superconductors \cite{AGD} and certain one-dimensional systems, 
in which the spectral gap is generated dynamically (for the latter
case a general proof is outlined in Ref. [\onlinecite{ET03}]).   

A metallic state with a small FS would necessarily be associated with
a Green's function that has both poles and zeroes at $\omega =0$. In
our previous work [\onlinecite{gf2}] we suggested a model for such a
state based on the quasi-one-dimensional Hubbard model at half
filling. The transverse hopping was treated in a Random Phase
approximation (RPA). In order to understand the conditions of stability of
such an exotic metal, one has to go beyond RPA, which is the main
subject of the present paper.
Experimental indications of the existence of SFS states come from ARPES
measurements in underdoped cuprates \cite{arpes} and from the Hall
effect measurements in heavy fermion materials \cite{Sielke}.


Before turning to the calculations, we shall give a qualitative account
of the physics we are after. Our starting point is 
an {\sl ensemble of uncoupled, Mott insulating chains}. The
relevant energy scale is the 1D Mott gap $m$. We consider finite
temperatures $T$ such that $T\ll m$. The physics is purely one
dimensional. 

We then turn on a small {\sl long range interchain tunneling}
with characteristic energy scale $t_\perp$. Clearly, at zero
temperature the hopping between chains will be essential and 
induce a 3D ordered state. On the other hand, in the window 
\be
t_\perp\ll T,\tilde{t}_\perp({\bf k}  )\ll m\ ,
\ee
we will recover the physics of 1D Mott insulating chains. Here
$\tilde{t}_\perp({\bf k})$ denotes the Fourier transform of the
interchain tunneling. Furthermore, as $T\ll m$ we may to a good
approximation work with zero temperature quantities in many
instances.

The crucial point is that while $t_\perp$ remains much smaller than the
Mott gap $m$, the Fourier transform $\tilde{t}_\perp({\bf k})$ can become
comparable to $m$ in some region of the Brillouin zone, i.e. we may
have a situation where 
\be
t_\perp\ll T\ll \tilde{t}({\bf 0})\approx m\ .
\label{window}
\ee

In this case an interesting {\sl ``hybrid'' state} combining 1D with
3D features develops. In particular, the low energy sector corresponds
to a 3D metal with a small Fermi surface and quasi-particles interacting
with well-developed collective modes. The existence of the regime 
(\ref{window}) is ensured by making the interchain tunneling long ranged.

\begin{figure}[ht]
\begin{center}
\epsfxsize=0.45\textwidth
\epsfbox{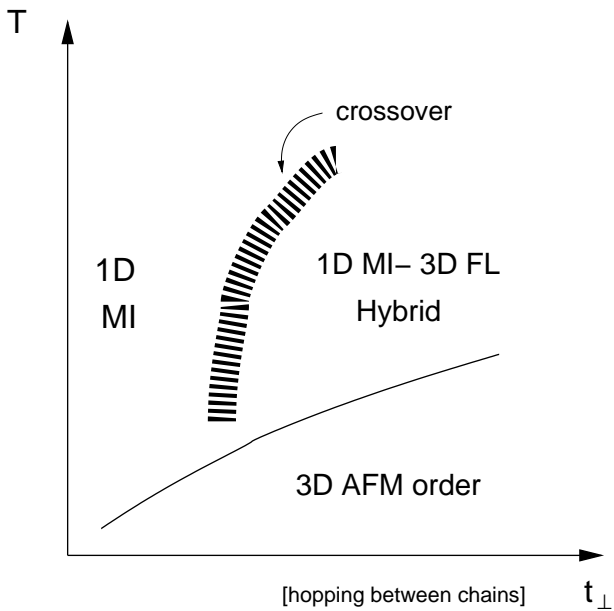}
\end{center}
\caption{Cartoon Phase Diagram for $T\ll m$ for weakly coupled 1D Mott
  insulators, where $m$ is the 1D Mott gap. }
\label{fig:MIFL}
\end{figure}

The dimensional crossover from a quasi one dimensional Mott insulator
to an anisotropic 3D Fermi liquid as a function of the strength
$t_\perp$ of the interchain hopping is sketched in Fig. \ref{fig:MIFL}.

The purpose of the present work is to derive an effective theory for
the low-energy degrees of freedom in the ``1D Mott insulator/3D Fermi
liquid hybrid'' regime and to analyze its instabilities towards 3D
order at sufficiently low temperatures.

\section{The model}
The model we study is the Hubbard model with a strongly anisotropic hopping:
\bea
H&=&-t\sum_{n,{\bf l},\sigma}\left[{c^{\dagger}_{n,{\bf l},\sigma}}
c_{n+1,{\bf l},\sigma}+{\rm h.c.}\right]
+U\sum_{n,{\bf l}} n_{j,{\bf l},\uparrow}n_{j,{\bf l},\downarrow}\nn
&&+ \sum_{{\bf l},{\bf m},n,p,\sigma}t_{{\bf
    l}{\bf m}}^{np}\ {c^{\dagger}_{n,{\bf l},\sigma}} c_{p,{\bf
m},\sigma}\ .
\label{Hamiltonian}
\eea
For definiteness we consider the chain direction to be $z$, so that
${\bf l}=(l_x,l_y), {\bf m}=(m_x,m_y)$ label Hubbard chains and
$n,p$ label the sites along a given chain. 

As we have mentioned before, the hopping integrals in the transverse
directions are supposed to be
small in comparison to $t$. In the limit $t_{\perp} =0$
and at half filling the model has a Mott-Hubbard gap $m$. We work in a
regime where the magnitude of this gap is much smaller than the
bandwidth $W \approx 4t$. In our previous paper \cite{gf2}
the transverse hopping was treated in a Random Phase approximation
(RPA). In order to suppress corrections to RPA (at least in some
temperature interval) we assume that the transverse hopping is long
ranged (see below).

\subsection{Uncoupled, Mott insulating chains}
Let us briefly discuss the low-energy physics for uncoupled chains. In
order to ease notations we suppress the chain index $({\bf l})$. 
Keeping only low-energy modes around the two Fermi points $\pm k_F$,
the electron annihilation operators are decomposed as 
\be
c_{n,\sigma}=\sqrt{a_0}\left[\exp(ik_Fx)\ R_\sigma(x)+
\exp(-ik_Fx)\ L_\sigma(x)\right],
\ee
where $a_0$ is the lattice spacing, $x=ja_0$ and $k_F=\pi/2a_0$.
The fermionic creation operators for left and right moving Fermions
are bosonized, using the following conventions
\bea
L^\dagger_\sigma(x)&=&\frac{\eta_\sigma}{\sqrt{2\pi}}e^{if_\sigma\pi/4}
\exp\left(-\frac{i}{2}\bar{\phi}_c\right)
\exp\left(-\frac{if_\sigma}{2}\bar{\phi}_s\right),\nn
R^\dagger_\sigma(x)&=&\frac{\eta_\sigma}{\sqrt{2\pi}}e^{if_\sigma\pi/4}
\exp\left(\frac{i}{2}\phi_c\right)
\exp\left(\frac{if_\sigma}{2}\phi_s\right),
\eea
where $\eta_a$ are Klein factors with $\{\eta_a,\eta_b\}=2\delta_{ab}$ and where
$f_\uparrow=1$, $f_\da=-1$.The chiral boson fields $\phi_a$ and
$\bar{\phi}_a$ fulfil the following commutation relations 
\be
[\phi_a(x),\bar{\phi}_a(y)]=2\pi i\ ,\quad a=c,s.
\ee
In terms of the chiral fields $\phi_a$ and $\bar{\phi}_a$ we define
canonical Bose fields $\Phi_a$ and their dual fields $\Theta_a$ by
\be
\Phi_a=\phi_a+\bar{\phi}_a\ ,\quad
\Theta_a=\phi_a-\bar{\phi}_a\ .
\ee
The low-energy effective Hamiltonian density for a single chain takes
the following bosonic form
\bea
{\cal H}_s &=& \frac{v_s}{16\pi}\left[(\p_x\Theta_s)^2 +
(\partial_x\Phi_s)^2\right]-g\ {\bf J}\cdot\bar{\bf J}, \label{spinboson}\nn 
{\cal H}_c &=& \frac{v_c}{16\pi}\left[(\p_x\Theta_c)^2 +
(\p_x\Phi_c)^2\right] +g\ {\bf I}\cdot\bar{\bf I}.
\label{SGM}
\eea
Here $I^\alpha$ and $\bar{I}^\alpha$ ($J^\alpha$ and $\bar{J}^\alpha$)
are the chiral components of the SU(2) pseudospin (spin) currents
\bea
I^z&=&-\frac{1}{4\pi}\partial_x\phi_c\ ,\quad
I^+=\frac{\eta_\up\eta_\da}{2\pi}\ e^{i\phi_c}\ ,\nn
J^z&=&-\frac{1}{4\pi}\partial_x\phi_s\ ,\quad
J^+=i\frac{\eta_\up\eta_\da}{2\pi}\ e^{i\phi_s}\ .
\eea
The current-current interaction in the spin sector of (\ref{SGM}) is
marginally irrelevant and we will ignore it in what follows. We note
that doing so enhances the symmetry in the spin sector from $SU(2)$
(spin rotational symmetry) to $SU(2)\times SU(2)$ (rotational symmetry
of the left and right sectors).

\subsubsection{Single-Particle Green's Function}

The single-particle Green's function for the half-filled Hubbard model
was obtained in the framework of the formfactor approach in Refs
[\onlinecite{gf2,LukZam01}]. In particular, when the charge and spin
velocities are equal we have 
\bea
G_0(\omega, \pm k_F+q) = \frac{Z_0}{\ri\omega \mp vq}\left[1 -
\frac{m}{\sqrt{m^2 + \omega^2 + (vq)^2}}\right]\!,\nn
\label{gt0}
\eea
where $Z_0 \approx 0.921862$. In order to obtain the above expression
for $G_0$ we took into account only processes involving the emission
of a single massive holon and a cascade of gapless spinons.

\subsubsection{Spin Sector}
The  spin operators
$S^\alpha_n=\frac{1}{2}c^\dagger_{n,a}\sigma^\alpha_{ab}c_{n,b}$ are expressed
in terms of the left and right moving Fermi fields by
\bea
S^\alpha_j &\simeq& (-1)^j\frac{a_0}{2}\left[R^\dagger_a(x)\
\sigma^\alpha_{ab}\ L_b(x)+{\rm h.c.}\right]\nn
&&+\frac{a_0}{2}\left[R^\dagger_a(x)\
\sigma^\alpha_{ab}\ R_b(x)+R\rightarrow L\right]\nn
&\equiv& a_0\left[(-1)^j n^\alpha(x)+J^\alpha(x)\right].
\eea
The bosonized expressions for the staggered components of the spin
operators are
\bea
R^\dagger_a(x)\ \sigma_{ab}^\alpha\ L_b(x)\simeq \frac{1}{\pi i\sqrt{2a_0}}
\exp\left(\frac{i}{2}\Phi_c\right)\ {\rm tr}\left(g\sigma^\alpha\right),
\eea
where we have replaced the product of Klein factors by their
expectation value
\be
\langle \eta_\up\eta_\da\rangle= -i\ ,
\ee
and where the matrix field $g$ is expressed in terms of the spin boson
$\Phi_s$ and its dual field $\Theta_s$ by
\bea
g = \sqrt{\frac{a_0}{2}}\left(
\begin{array}{cc}
\exp\left(\frac{i}{2}\Phi_s\right) & i\exp\left(-\frac{i}{2}\Theta_s\right)\\
i\exp\left(\frac{i}{2}\Theta_s\right) & \exp\left(-\frac{i}{2}\Phi_s\right)
\end{array}\right).
\label{gfab}
\eea
 At $T =0$ we have 
\bea
\la g_{\alpha\beta}(\tau,x) g_{\gamma\delta}^\dagger(0,0)\ra  =
 \delta_{\alpha\delta}\delta_{\beta\gamma}\frac{a_0}{2\sqrt{v^2\tau^2 +
 x^2}}.
\label{prop} 
\eea
Using (\ref{gfab}) one can easily calculate multi-point correlation
functions of $g$. 

The action (\ref{spinboson}) describing the collective spin
excitations on each chain is equivalent to the SU$_1$(2)
Wess-Zumino-Novikov-Witten (WZNW) model once we drop the marginally
irrelevant interaction of spin currents. The WZWN action for the
matrix field $g(\tau,x)$ is given by
\bea
W[g]&=&\frac{1}{16\pi}\int d^2x\ {\rm Tr}
\left[\partial^\mu g^\dagger\partial_\mu  g\right]\nn
&&+\frac{\epsilon_{\mu\nu\lambda}}{24\pi}\int_B d^3x\
{\rm   Tr}\left[g^\dagger\partial^{\mu}g g^\dagger\partial^{\nu}g
g^\dagger\partial^{\lambda}g\right],
\label{wzw}
\eea
where $x_1=v\tau$, $x_2=x$, $\partial_\mu=\frac{\partial}{\partial x^\mu}$,
$B$ is a three dimensional half-space ($x_3\leq 0$)
whose boundary at $x_3=0$ conincides
with the two-dimensional $(v\tau,x)$-plane and $g$ is an arbitrary
extrapolation of the function defined on the two-dimensional space $x_3=0$,
which approaches $1$ at $x_3\to-\infty$. The action (\ref{wzw}) is
invariant under $SU(2)\times SU(2)$ transformations 
$g\rightarrow Ug\widetilde{U}^\dagger$. The marginally irrelevant
interactions of spin currents breaks this symmetry down to the
diagonal spin-rotational $SU(2)$ $g\rightarrow UgU^\dagger$.
The form (\ref{wzw}) of the action for the spin degrees of freedom is
significantly more complicated than the free boson representation
(\ref{spinboson}). The latter is very convenient for calculations in one
dimension, but may be less useful when one considers interchain
coupling due to the fact that the dual field $\Theta_s$ is non-local
with respect to $\Phi_s$. The formulation in terms of the matrix field $g$
has the advantage of the fundamental field being the order parameter
itself. In fact, $W[g]$ is the Ginzburg-Landau action for a 1D
spin-1/2 antiferromagnet.  

\subsubsection{Three Point Function}
\label{3-point}

\begin{figure}[ht]
\begin{center}
\epsfxsize=0.4\textwidth
\epsfbox{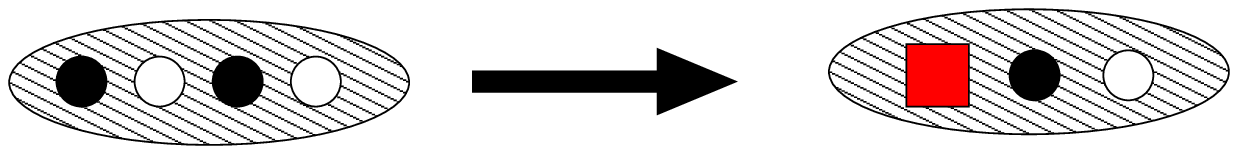}
\end{center}
\caption{The fermion-spinon interaction generated by fusion
(\ref{single}). }
\label{vertex}
\end{figure}

An important ingredient in our analysis are three-point functions of
the form $\langle \tr[g({\bf z})\sigma^\alpha]\ R^\dagger_a({\bf {\bf
z_1}})\ L_b({\bf {\bf z_2}})\rangle$. The large-distance asymptotics
of these correlators can be evaluated by using the results of
\cite{LukZam01} 
\begin{widetext}
\bea
\langle \tr[g({\bf z})\sigma^+]\ R^\dagger_\da({\bf {\bf z_1}})\ 
L_\up({\bf {\bf z_2}})\rangle
&=&-i\frac{\langle\eta_\da\eta_\up\rangle}{2\pi}
\langle \tr[g({\bf z})\sigma^+]\ e^{-\frac{i}{2}\phi_s({\bf z_1})}\ 
e^{\frac{i}{2}\bar{\phi}_s({\bf z_2})}\rangle_s
\langle e^{\frac{i}{2}\phi_c({\bf z_1})}\ e^{\frac{i}{2}\bar{\phi}_c({\bf z_2})}\rangle_c\nn
&\simeq& i\frac{\langle\eta_\da\eta_\up\rangle}{2\pi}
\langle \tr[g({\bf z})\sigma^+]\ e^{-\frac{i}{2}\phi_s({\bf z_1})}\ 
e^{\frac{i}{2}\bar{\phi}_s({\bf z_2})}\rangle_s
\frac{Z_1\sqrt{a_0}}{\pi}e^{\frac{3i\pi}{4}}\ K_0(mr_{12})\nn
&\simeq& \widehat{Z}\langle\eta_\da\eta_\up\rangle\
K_0(mr_{12})
\langle \tr[g({\bf z})\sigma^+]\ \tr[g({\bf z_+})\sigma^-]\rangle_s,\nn
\label{3point}
\eea
where ${\bf z_{1,2}}=(\tau_{1,2},x_{1,2})$, $r_{12}=|{\bf z_1}-{\bf
  z_2}|$ and ${\bf
  z_+}=(\frac{\tau_1+\tau_2}{2},\frac{x_1+x_2}{2})$. The constant
$\widehat{Z}$ is related to the normalisation $Z_0$ of the single-particle
Green's function by \cite{LukZam01}
\end{widetext}
\be
\widehat{Z}=-\frac{Z_0}{\pi^\frac{3}{2}}\sqrt{\frac{m}{va_0}}.
\ee
The calculation we have just carried out can be summarized by the
following approximate relations
\bea
R^\dagger_a({\bf z_1})\sigma^\alpha_{ab}L_b({\bf  z_2})
&\longrightarrow&{i\widehat{Z}}\ K_0(mr_{12})
  \ {\tr}[g({\bf z_+})\sigma^\alpha]\ ,\nn
L^\dagger_a({\bf z_1})\sigma^\alpha_{ab}R_b({\bf  z_2})
&\longrightarrow&{i\widehat{Z}}\ K_0(mr_{12})
  \ {\tr}[g({\bf z_+})\sigma^\alpha]\ .\nn
\label{single}
\eea
The approximation (\ref{single}) fails at small distances.
In order to remove the logarithmic singularity of $K_0$ one needs to
include terms corresponding to the multiple production of solitons and
antisolitons. At energies much smaller than the Mott gap, the fusion 
(\ref{single}) gives rise to the spin-fermion vertex depicted on
Fig. \ref{vertex}. 
 
\subsection{Long range interchain hopping}

In order to have a small parameter in our theory we consider the
interchain hopping to be long-ranged, such that the Fourier transform
of the hopping matrix elements strongly depends on the wave
vector. This is a well-known trick (see, for example
Ref. [\onlinecite{gorkov}]) and results in a controlled ``loop''
expansion, where every integration over the transverse momenta leads
to a small factor $\kappa_0^2$ in three dimensions, where $\kappa_0$
is the inverse range of the interchain tunnelling. The interchain
hopping may be taken long ranged both along and perpendicular to the
chain direction. In order to simplify the calculations, we will
constrain our discussion to the case where 
$t_{{\bf l}{\bf m}}^{np}=t({\bf l}-{\bf m})\delta_{n,p}$, 
i.e. the interchain hopping has no component along the chain direction
and depends only on the distance between chains. The Fourier transform
of the interchain tunneling is then given by 
\be
\tilde{t}_{\perp}({\bf k}_\perp)=\sum_{{\bf m}}
t_{{\bf l}{\bf m}}\ \exp(i{\bf k}_\perp\cdot (\bl-\bm)a_\perp).
\ee
In the following we choose the interchain hopping such that it respects the
particle-hole symmetry 
\bea
c_{n,{\bf l},\sigma}\longleftrightarrow (-1)^{n+l_x+l_y}
c_{n,{\bf l},\sigma}^{\dagger}\ ,
\eea
which implies that
\be
\tp({\bf k}_\perp + {\bf Q}) = - \tp({\bf k}_\perp),
\ee
where 
\be
{\bf Q} = \left(\frac{\pi}{a_\perp},\frac{\pi}{a_\perp}\right)
\ee
is the antiferromagnetic wave vector in the direction transverse to
the chains. It is straightforward to generalize our following analysis
to non particle-hole symmetric cases. The basic assumptions underlying
our model are then summarized in the following inequalities: 
\bea
W \gg m\sim |\tilde{t}_{\perp}({\bf 0})|=|\tilde{t}_{\perp}({\bf Q})|  
\gg
\tilde{t}_{\perp}({\bf p}_\perp)\ .
\label{assump}
\eea
Here $W=4t$ and $m$ are the band width and Mott gap for uncoupled chains
respectively and $|{\bf p}_\perp a_\perp|,|({\bf p}_\perp-{\bf Q}) a_\perp|
\gg \kappa_0$. The small parameter $\kappa_0$ characterizes
the support of $\tilde{t}_\perp({\bf  k}_\perp)$ in momentum space. 
The precise form of the momentum dependence of $\tp$ is supposedly
unimportant, but in order to simplify the concrete calculations we
shall use the following model: 
\bea
\tp({\bf k}_\perp) & =& 
 -\frac{t_0}{1 + |{\bf k}_\perp a_\perp|^2
\kappa_0^{-2}}, ~~
|{\bf k}_{\perp}a_\perp| \ll 1.
\label{dependence}
\eea
Within the model (\ref{dependence}) the integration over the transverse
wave vectors may be replaced by integration over $t \equiv
t_{\perp}({\bf k}_\perp)$
\bea
a_\perp^2\int\frac{d^2k_\perp}{4\pi^2}
f(t)\approx
\frac{\kappa_0^2t_0}{4\pi}\int_{\frac{\kappa_0^2t_0}{4\pi^2}}^{t_0}
\frac{dt}{t^2}\left[f(t)+f(-t)\right].
\label{dos1} 
\eea
Some readers may find that our approach is similar to Dynamical
Mean Field theory in an {\it infinitely dimensional} space. This is not
the case: the difference comes from the fact that in our model the
transverse density of states is constant on the zone boundary. This
feature strengthens the influence of fermionic coherent modes and
utterly changes the physics (see the discussion in the Conclusions).

\section{Perturbation Theory in the interchain tunneling}

As we have already mentioned, the perturbation theory in the
interchain tunneling can be reorganized in terms of a loop
expansion. Every integration over the transverse momenta generates a
small factor $\kappa_0^2$. 
We will refer to the leading order ${\cal O}(\kappa_0^0)$ in this
expansion as ``Random Phase Approximation'' (RPA).

\begin{figure}[ht]
\begin{center}
\epsfxsize=0.4\textwidth
\epsfbox{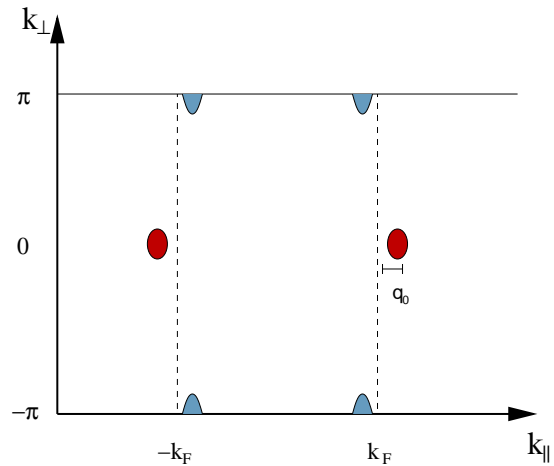}
\end{center}
\caption {The Brillouin zone with the electron (red ovals) and
  hole-like (blue semi-ovals) Fermi pockets of a two-dimensional
lattice. The noninteracting Fermi surface is shown as a dashed line.}
\label{fig:surface}
\end{figure}

The RPA expression for the single particle Green's function $G$ was
derived in Ref. [\onlinecite{gf2}] and is given by
\bea
G(\omega,\pm k_F+q, {\bf k}_\perp) = 
\frac{G_0(\omega,\pm k_F+ q)}{1-\tilde{t}_{\perp}({\bf k}_\perp)
G_0(\omega,\pm k_F+q)}\ .
\label{G2}
\eea
Here $G_0$ is the single-particle Green's function for an individual
chain (\ref{gt0}). In a purely one-dimensional Mott insulator the electron
is a composite particle and as a result the spectral function is
incoherent. Coherent electronic excitations reappear as soon as the
interchain tunneling is turned on. They can be understood as
antiholon-spinon bound states and occur at energies {\sl below} the
Mott gap. When $t_{\perp}$ exceeds a certain critical value, the
dispersion of the coherent mode crosses the chemical potential and a
Fermi surface is formed. As a consequence of particle-hole symmetry,
at half-filling this Fermi surface consists of electron- and hole-like
pockets of equal volume. A sketch of such a Fermi surface is shown in
Fig. \ref{fig:surface}.
A convenient measure for the strength of the interchain coupling is
given by the quantity
\be
{\cal Z}\equiv\frac{Z_0t_0}{m}\ ,
\label{Z}
\ee
where $t_0$ is defined in (\ref{dependence}). The RPA form 
(\ref{G2}) for the
Green's function features a pole corresponding to a coherent
quasi-particle mode. This mode crosses the chemical potential when
${\cal Z}$ exceeds the critical value 
\be
{\cal Z}_c=3.33019\ldots\ ,
\label{Zc}
\ee
and a Fermi surface is present for all ${\cal Z}>{\cal Z}_c$.

\begin{figure}[ht]
\begin{center}
\epsfxsize=0.4\textwidth
\epsfbox{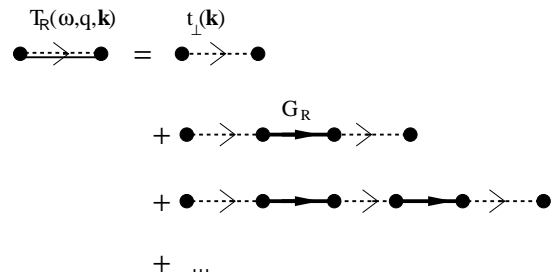}
\end{center}
\caption{The dressed interchain hopping.}
\label{fig:TR}
\end{figure}

Having in hand the expression for the chain single-particle Green's
function, we may use it to define a dressed interchain hopping
$\tilde{T}_{R,L}(\omega,q,{\bf k}_\perp)$ by summing the diagrams shown in
Fig. \ref{fig:TR}.
This results in
\bea
T_{R,L}(\omega,q,{\bf k}_\perp)&=&\frac{\tilde{t}_\perp({\bf
    k}_\perp)}{1-{\tilde t}({\bf k}_\perp)\ G_0(\omega,\pm k_F+q)}\ ,
\eea
where the $+$ sign corresponds to $R$ and the $-$ sign to $L$
respectively. We note that the dressed interchain hopping is equal
to the propagator of the Hubbard-Stratonovich field that can be
introduced to decouple the interchain hopping \cite{Boies}.

\subsection{The Spin Sector in RPA}

\begin{figure}[ht]
\begin{center}
\epsfxsize=0.45\textwidth
\epsfbox{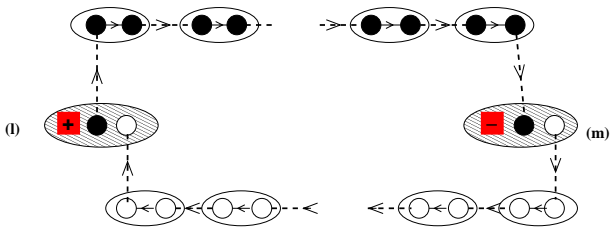}
\end{center}
\caption{Real-space diagrams that contribute to the two-point function of
  staggered magnetizations between chains $(l)$ and $(m)$.} 
\label{fig:spindiag}
\end{figure}

In the RPA, the spin sector remains one-dimensional and critical. This
can be seen as follows. Let us consider the real-space correlator
between the staggered components of spins on different chains ${\bf
  l}$ and ${\bf m}$
\bea
\langle n^{+}_{j,\bl}(t)\  n^{-}_{1,\bm}(0)\rangle.
\eea
Within perturbation theory in the interchain hopping, we need at least
one right moving and left moving fermion operator each on chains $l$
and $m$ in order to obtain a nonzero expectation value in the spin sector.
The only ways to achieve this are shown in Fig.\ref{fig:spindiag}.
Here the dashed lines denote the bare interchain hopping, ellipses
enclosing two black (white) circles represent the purely 1D Green's
function of right (left) moving electrons on a given chain and the
ellipses enclosing two circles and a hexagon stand for the three-point
function (\ref{3point}).
Clearly all such diagrams involve at least one integration over the
transverse momentum. Hence, within the RPA spin-spin correlation
functions remain entirely one-dimensional and spins on different
chains remain uncorrelated.

\subsection{Excitation Spectrum in RPA}
From the discussion above it is clear that for sufficiently strong
interchain hopping ${\cal Z}>{\cal Z}_c$ the RPA leads to two types of
gapless excitations
\begin{itemize}
\item{}Fermionic particle and hole excitations over the Fermi surface
with anisotropic 3D dispersions.
\item{}Collective excitations of the spin degrees of freedom. These
are of a purely 1D nature and do not have a dispersion in
the direction transverse to the chains.
\end{itemize}
If one goes beyond the RPA, interactions between these two types of
excitations will be generated. In the following we determine the form
of these interactions and study their effects. To go beyond the
RPA in principle requires the knowledge of the two-particle Green's
function of uncoupled Mott-insulating chains. However, if one
restricts one's attention to the regime of energies small compared to
the Mott gap, the three-point function \r{3point} (which corresponds to a
particular limit of the two-particle Green's function) suffices.
\section{Interchain Exchange and Estimate of the Transition
  Temperature}
\label{sec:exch}
\begin{figure}[ht]
\begin{center}
\epsfxsize=0.25\textwidth
\epsfbox{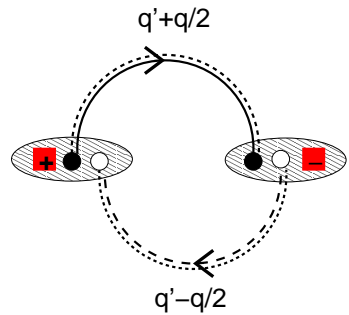}
\end{center}
\caption{Leading order ${\cal O}(\kappa_0^2)$ contribution to the
  interchain exchange.} 
\label{fig:J}
\end{figure}

Although it is obvious that corrections to RPA are of higher order in
the small parameter $\kappa_0$, they will diverge at small
temperatures. Therefore RPA works only at finite temperatures and for 
its consistency the transition temperature (below which the system is
three-dimensionally ordered) must be much smaller than the
Mott-Hubbard gap $m$. It is therefore important to estimate the
corresponding corrections to RPA and their temperature dependence. 
As the first step in taking  into account corrections to RPA we have
to estimate the interchain RKKY interaction. As we have shown in
section \r{3-point}, there is a three-point ``vertex'' that couples
the spin degrees of freedom to the fermionic quasi-particles. In second
order perturbation theory in this interaction, an interchain exchange
interaction between the spin degrees of freedom is generated. The
corresponding action is given by
\bea
S_{\rm xc}&=&\int\prod_{j=1}^2 d\tau_j dx_j\sum_{\bl\neq \bm}
J_{\bl\bm}(x_1-x_2,\tau_1-\tau_2)\nn
&&\qquad\times\ \tr\left[g_\bl(\tau_1,x_1)\
  g^{\dagger}_\bm(\tau_2,x_2)\right]. 
\label{Jint}
\eea
The Fourier transform of the leading order (in $\kappa_0$) exchange
matrix element is given by the ``bubble'' diagram shown in
Fig. \ref{fig:J}, where the doubles lines are the dressed interchain
hoppings for left and right moving fermions and the squares denote
the elements of the matrix field.
The result is
\begin{widetext}
\bea
\tilde{J}(\omega,q,{\bf q_\perp})&=&\widehat{Z}^2
\int\frac{d\omega^\prime dq^\prime}{(2\pi)^2}
a_\perp^2\int{d^2k_\perp}{}
\left[\frac{v}{m^2+\omega'^2+(vq')^2}\right]^2\
{T}_R\left(\omega^\prime+\frac{\omega}{2},q'+\frac{q}{2},{\bf
  k_\perp}\right)\nn
&&\qquad\qquad\qquad\times {T}_L\left(\omega^\prime-\frac{\omega}{2},q'-\frac{q}{2},{\bf
  q_\perp-k_\perp}\right)\Bigr].
\label{xch}
\eea
\end{widetext}
 To make the calculations easier, we use the $k$-dependence
(\ref{dependence}) so that the sum over the transverse wave
vectors is replaced by integration over $t$ according to (\ref{dos1}). 
Since $\tilde{t}({\bf k}_\perp)$ is peaked near zero and ${\bf Q}$, there
are two interesting wave vectors: ${\bf q_\perp}=0$ and ${\bf
  q}_{\perp} = {\bf Q}$. 
\subsection{Case 1: ${\cal Z}<{\cal Z}_c$}
In this case the coherent electron modes still have a gap and no Fermi
surface is formed in the RPA. Using (\ref{dos1})
to carry out the summation over
transverse momenta we obtain
\bea
\tilde{J}(0,0,{\bf 0}) \approx{\cal C}_1
\int_0^\infty ds\ s\frac{\arctan(\z0 G(s))}{(1+s^2)^2\ G(s)}\equiv
\alpha_0{\cal C}_1,
\label{J0}
\eea
where ${\cal Z}=Z_0t_0/m$ is defined in (\ref{Z}) and 
\be
{\cal C}_1=\zh^2{\kappa_0^2t_0}{}
\frac{v}{mZ_0}=\frac{{\cal Z}\kappa_0^2}{\pi^3a_0}m\ ,
\ee
and
\bea
G(s) = s^{-1}\left[1 - \frac{1}{\sqrt{s^2 + 1}}\right].
\eea
As expected, this interaction is of order of $\kappa_0^2 t_0$. The
numerical factor $\alpha_0$ ranges between $0$ for ${\cal Z}=0$ and
$2.81$ for ${\cal Z}\to 3.33019$. The exchange at momentum transfer
${\bf Q}$ is 
\bea
\tilde{J}(0,0,{\bf Q}) &\approx&-\frac{{\cal C}_1}{2}
\int_0^\infty ds\ s
\frac{\ln\left[\frac{1+\z0 G(s)}{1-\z0 G(s)}\right]}{(1+s^2)^2\
G(s)}\nn
&=&-\alpha_Q{\cal C}_1,
\label{JQ}
\eea
where $\alpha_Q$ varies between $0$ for for ${\cal Z}=0$ and
$3.07$ for ${\cal Z}\to {\cal Z}_c$. 

\subsection{Case 2: ${\cal Z}>{\cal Z}_c$}
\label{ssec:case2}
In this case a Fermi surface in the form of electron and hole pockets
is present. The presence of zero energy modes does not really affect
the exchange at zero momentum transfer, which is given by
\bea
\tilde{J}(0,0,{\bf 0}) \approx {\cal C}_1
\int_0^\infty ds\ \frac{sf(s)}{(1+s^2)^2\  G(s)}
=\alpha'_0{\cal C}_1\ ,
\label{J02}
\eea
where
\bea
f(s)&=&2\arctan(\xi(s)\z0 G(s))-\arctan(\z0 G(s))\ ,\nn
\xi(s)&=&{\rm min}(1,[{\cal Z}G(s)]^{-1})\ .
\eea
We find that $\alpha'_0$ starts at $1.405$ for $\z0\to 3.33019$, then
goes through a maximum of approximately $1.48$ around $\z0\approx 4.18$
and then diminishes slowly. Hence the exchange at wave number zero
play a subdominant role. 

Let us now turn to the exchange at wavenumber ${\bf Q}$. We find that
there is a logarithmic divergence in (\ref{xch}), which is related
to the Fermi surface formation and will be discussed later in more
detail. We regularize the divergence by temperature. This may be done
by replacing the $\omega'$-integral in (\ref{xch}) by a sum over Matsubara
frequencies and substituting the finite-temperature Green's function
for their $T=0$ analogs in the dressed interchain hoppings
$\tilde{T}_{R,L}$ in (\ref{xch}). At low temperatures the single particle
Green's function is given by \cite{ET03}
\bea
&&G_R(\omega_n,q)=\int_{-\infty}^\infty dx\frac{A_R(x,q)}{i\omega_n-x}\
,\nn
&&A_R(\omega,q)=\frac{Z_0}{4\pi^2  m}
\sqrt{\frac{2m}{T}}\left[\frac{m}{|\omega-vq|}\right]^{\frac{3}{2}}\nn
&&\quad\times {\rm Re}\left[\sqrt{-2i}
  B\left(\frac{1}{4}-\frac{i}{2}\frac{\omega^2-v^2q^2-m^2}{2\pi
    T|\omega-vq|},\frac{1}{2}\right)\right].
\label{finitetgf}
\eea
The singular piece of $\tilde{J}^{\rm sing}(0,0,{\bf Q})$
diverges logarithmically with temperature and is estimated as
\bea
\tilde{J}^{\rm sing}(0,0,{\bf Q}) \approx -{\cal C}_1\
\ln\left[\frac{m}{T}\right] \int_{s_-}^{s_+}ds\
\frac{s[G(s)]^{-1}}{(1+s^2)^2} ,
\label{jsing}
\eea
where $s_\pm$  are solutions to the equation
\bea
1-\z0 G(s_\pm)=0\ .
\eea
In order to establish the exchange at nonzero values of $\omega$ and
$q$ we have calculated $\tilde{J}(\omega,q,{\bf Q})$ numerically at
small temperatures. Rather than using the finite-$T$ Green's function
(\ref{finitetgf}) we work with the $T=0$ expression (\ref{gt0}) and
replace $\omega$ by the discrete Matsubara frequencies. For small
temperatures this is a reasonable approximation. We find that
$|\tilde{J}(\omega,q,{\bf Q})|$ is largest at $\omega=0=q$.

In addition to the singular piece (\ref{jsing}) there also is a
regular contribution to the exchange. As long as we are close to the
transition, i.e.
\be
\frac{{\cal Z}-{\cal Z}_c}{{\cal Z}_c}\ll 1\ ,
\label{region}
\ee
we may estimate the regular contribution to the exchange by its value
at the critical strength ${\cal Z}_c$ of the interchain tunneling. The
latter is given by (\ref{JQ})
\bea
\tilde{J}^{\rm reg}(0,0,{\bf Q}) \approx -3.071\ {\cal C}_1.
\label{jreg}
\eea
In the regime (\ref{region})
\be
\tilde{J}^{\rm sing}(0,0,{\bf Q}) \approx -1.318\
\sqrt{{\cal Z}-{\cal Z}_c}\
{\cal C}_1\ \ln\left[\frac{m}{T}\right] .
\label{Jsing}
\ee
The total exchange constant at wave number ${\bf Q}$ is then estimated
as 
\bea
\tilde{J}(0,0,{\bf Q}) \approx -{\cal C}_1\left[3.071+
1.318\
\sqrt{{\cal Z}-{\cal Z}_c}\ \ln\left[\frac{m}{T}\right]\right].
\label{jtot}
\eea

Having determined the exchange constant we are now in a position to
estimate the temperature at which a magnetic instability develops.
In the absence of interchain hopping the correlation functions of the 
matrix field at $T=0$ are given by (\ref{prop}).
At $T>0$ we have
\bea
\langle g_{\alpha\beta}(\tau,x) g^+_{\gamma\delta}(0,0)\rangle = 
\delta_{\alpha\delta}\delta_{\beta\gamma}
\frac{\pi Ta_0/v}{2|\sinh\left(\frac{\pi T}{v}[x+iv\tau]\right)|},
\label{correlT}
\eea
if we neglect the marginally irrelevant current-current interaction in
the spin-sector of the Hamiltonian (\ref{SGM}) describing the 1D Mott
insulating chains. If one takes it into account in renormalization group
improved perturbation theory one obtains \cite{Barzykin00a}
\be
\langle {\rm tr}(g\sigma^\alpha)\ {\rm
  tr}(g^\dagger\sigma^\beta)\rangle=
\delta_{\alpha\beta}
\frac{\sqrt{\ln\left[\frac{\Lambda}{T}\right]}\ \pi Ta_0v^{-1}}
{|\sinh\left(\frac{\pi T}{v}[x+iv\tau]\right)|},
\ee
where $\Lambda$ is a high-energy cutoff, which we may take to be of
the order of the hopping integral along the chain direction.
Carrying out an analogous calculation for the dimerization operator we find
\be
\langle {\rm tr}(g)\ {\rm tr}(g^\dagger)\rangle=
\delta_{\alpha\beta}
\frac{\left[\ln\left[\frac{\Lambda}{T}\right]\right]^{-\frac{3}{2}}\ \pi Ta_0v^{-1}}
{|\sinh\left(\frac{\pi T}{v}[x+iv\tau]\right)|}.
\label{dimer}
\ee
Upon Fourier transformation and analytical contibuation one finds the
following result for the dynamical magnetic susceptibility of the
uncoupled chain system \cite{sb83} 
\bea
\chi_{\rm 1d}(\omega, q) &=& -
\frac{a_0\sqrt{\ln\left[\frac{\Lambda}{T}\right]}
}{2T}
\frac{\Gamma\left[\frac{1}{4}- i\frac{\omega - vq}{4\pi T}\right]
\Gamma\left[\frac{1}{4}- i\frac{\omega + vq}{4\pi T}\right]}
{\Gamma\left[\frac{3}{4}- i\frac{\omega - vq}{4\pi T}\right]
\Gamma\left[\frac{3}{4}- i\frac{\omega + vq}{4\pi T}\right]}.\nn
\label{susco}
\eea
The dimerization susceptibility is equal to (\ref{susco}) apart from the
prefactor, in which $\sqrt{\ln[\Lambda/T]}$ is
replaced by $\left[\ln\left[\Lambda/T\right]\right]^{-3/2}$. Hence the
staggered spin susceptibility is always more singular than the
dimerization susceptibility and as a result the dominant instability
of the spin sector is towards N\'eel order. The enhancement of the
spin susceptibility as compared to the dimerization susceptibility is
caused by the leading irrelevant operator in the Hamiltonian, namely
the interaction of spin currents. If we were to add an interaction to
the underlying lattice Hamiltonian in order to eliminate this
interaction, the symmetry between dimerization and the staggered
components of the spins would be broken by some other irrelevant operator.
The the dynamical susceptibility of the coupled chains system can be
determined by a expansion in the interchain coupling of the type
discussed in \cite{marc,katanin}. The leading term is given by the
Random Phase Approximation
\bea
\chi_{\rm 3d}(\omega, q,{\bf p}_\perp) &=& \frac{\chi_{\rm 1d}(\omega,q)}
{1-2\tilde{J}(\omega,q,{\bf p}_\perp)\chi_{\rm 1d}(\omega,q)}.
\label{chiRPA}
\eea
Given the expression (\ref{chiRPA}) for the dynamical susceptibility we
may obtain an estimate for the transition temperature $T_c$ below
which three dimensional magnetic long range order develops. $T_c$ is
defined as the temperature at which a zero frequency pole develops in
$\chi_{\rm 3d}$. Given that $\chi_{\rm 1d}(0,q)$ is peaked at $q=0$ and
$\tilde{J}$ is peaked at $q=0$ and ${\bf p}={\bf Q}$ we obtain the
following condition fixing $T_c$
\bea
1 - 2\tilde{J}(0,0,{\bf Q})\chi_{\rm 1d}(0,0) =0.
\eea 
Replacing $\tilde{J}(0,0,{\bf Q})$ by (\ref{jtot}) we arrive at the
following equation determining the transition temperature $T_c$ 
\bea
\frac{T_c}{m}&\approx& 2.887\kappa_0^2
\sqrt{\ln\left[\frac{\Lambda}{T_c}\right]}\nn
&&\times \left[1+0.429\sqrt{{\cal Z}-{\cal Z}_c}\
\ln\left[\frac{m}{T_c}\right]\right].
\label{tc1}
\eea
Let us consider the two limiting cases in which either the regular
(\ref{jreg}) or the singular part (\ref{Jsing}) of the exchange dominates and
drives the transition. The first case occurs if we are very close to
the point where the Fermi pockets are first formed and ${\cal Z}-{\cal
  Z}_c\ll(\ln(\kappa_0^2))^{-2}$. Then the transition temperature is
roughly equal to
\be
\frac{T_c}{m}\approx 2.887\ \kappa_0^2
\sqrt{\ln\left[\frac{\Lambda}{m\kappa_0^2}\right]}. 
\ee
The second case occurs if ${\cal Z}-{\cal
  Z}_c\gg(\ln(\kappa_0^2))^{-2}$ and then
\be
\frac{T_c}{m}\approx 1.239\ \kappa_0^2\delta
\sqrt{\ln\left[\frac{\Lambda}{m\kappa_0^2\delta}\right]}
\ln\left[\frac{1}{\kappa_0^2\delta}\right],
\label{tc2}
\ee
where $\delta=\sqrt{{\cal Z}-{\cal Z}_c}$.

\section{Effective Theory at Low Energies; the Residual Interactions}

Now we are in position of writing the low-energy effective action for
the metallic state. This effective action describes interactions of the
low-energy modes i.e. the coherent fermions and the order parameter
field $g_{ab}$. In the previous section we calculated the interchain
coupling for the $g$-field.  It contains a part coming from states far
from the chemical potential and the part with logarithmic divergences
coming from the states close to the Fermi surface. We can isolate the
first piece  and include it into the effective action of $g$
\bea
S_{\rm sp} &=& \sum_{\bf n} W[g_{\bf n}]\nn
&+&\sum_{\bm,\bl}J_{\bm\bl}\int\! d\tau\ \! dx
\tr\left[g_{\bm}(\tau,x)g_\bl^{\dagger}(\tau,x)\right],
\label{spin}
\eea
where to first approximation 
\bea
J_{\bn\bl} \approx \frac{Z_0^2}{\pi^2 ma_0}t^2_{\bn\bl}\ .
\eea
This part of the action plays the role of the sigma model in the
spin-fermion model by Pines and Chubukov (see \cite{chub2} and
references therein). Taken in isolation, this model has an instability
at some temperature $T_c$ which can be estimated from the RPA
expression for the dynamical magnetic susceptibility in complete
analogy with our calculation in section \ref{ssec:case2}.
Since the coherent fermions are low-energy excitations, they cannot be
simply integrated out, but their interaction with  $g$-field should
be added to the action. The Fermi surface of the coherent fermions is
determined by the equation 
\bea
G_0(0,q)\ \tilde{t}_{\perp}({\bf k}_\perp) = 1,
\eea
and consists of four pockets (two electron-like ones and two hole-like
ones) as is shown in Fig.\ref{fig:surface}. Let us consider the
situation where the scale of the interchain hopping $t_0$ is very
slightly larger than the minimal value $\bar{t}_0$ required for
the formation of a Fermi surface
\be
t_0=\tm+\dt=\sqrt{\frac{11+5\sqrt{5}}{2}}\frac{m}{Z_0}+\dt\ .
\ee
The electron and hole pockets are then shallow and anisotropic and the Fermi
surface is determined by the equation
\be
E(q,{\bf k}_\perp)=0\ ,
\label{e=0}
\ee
where
\bea
E(q,{\bf k}_\perp)&=&A_\parallel m\frac{(q-q_0)^2}{q_0^2}+A_\perp m
\frac{|{\bf k}_\perp a_\perp|^2}{\kappa_0^2}-E_0\ ,\label{E}\\ 
E_0&\approx&0.352 \delta t\ ,\quad
\frac{vq_0}{m}=\left[\frac{1+\sqrt{5}}{2}\right]^\frac{1}{2}\approx 1.27202,\nn
A_\parallel&\approx& 0.543\ ,\quad
A_\perp\approx 1.27202\ .
\eea
The electron pockets are formed at $(k_F+q,{\bf k}_\perp)$ and $(-k_F-q,{\bf
  k}_\perp)$ whereas the hole pockets are located at $(k_F-q,{\bf
  Q}+{\bf k}_\perp)$ and $(-k_F+q,{\bf Q}+{\bf k}_\perp)$, where $q$
and ${\bf k}_\perp$ are determined from (\ref{e=0}). Let us
denote the annihilation operator of the coherent fermions by
$\Psi(\tau,q,{\bf k}_\perp)$. The soft modes occur in the vicinity of the
electron  and hole pockets and it is convenient to decompose
$\Psi(\tau,q,{\bf k}_\perp)$ accordingly. We denote by $R_e(\tau,q,{\bf p}_\perp)$
and $L_e(\tau,q,{\bf p}_\perp)$ the annihilation operator in the vicinity of
the electron pockets and $(q,{\bf p}_\perp)$ is the deviation from $(\pm k_F,{\bf
  0})$. Similarly we denote by $R_h(\tau,q,{\bf p}_\perp)$ and
$L_h(\tau,q,{\bf p}_\perp)$ the annihilation operator in the vicinity of the
hole pockets and $(q,{\bf p}_\perp)$ is the deviation from $(\pm k_F,{\bf Q})$.

From Eq.(\ref{E}) we determine the particle density associated with a
single pocket is 
\bea
n \approx 0.027 a_{\perp}^{-2} q_0\kappa_0^2({\cal Z} - {\cal Z}_c)^{3/2}
\eea
The liquid of quasi-particles becomes degenerate at temperatures of
order of $E_0$. Comparing  it with the transition temperatures
(\ref{tc1}), (\ref{tc2}) we conclude that the degenerate metallic state
exists only at  
\bea
{\cal Z}-{\cal Z}_c \gg \kappa_0^2\ .
\eea
corresponding  to $na_{\perp}^2 \gg 0.027 q_0\kappa_0^5$.
Close to the Fermi surface the Green's function (\ref{G2}) can  be
approximated as   
\bea
G(\omega,\pm k_F+q,{\bf k}_\perp) &\approx& \frac{Z_2}{i\omega - E(\pm
  q,{\bf k}_\perp)}\ ,\\
G(\omega,\pm k_F+ q,{\bf k}_\perp +{\bf Q}) &\approx& \frac{Z_2}
{i\omega + E(\mp q,{\bf k}_\perp)}\ ,
\label{Geh}
\eea
where
\be
Z_2\approx\frac{vq_0}{\tm} \approx 0.352\ .
\ee
The expressions (\ref{Geh}) exhibit the particle-hole symmetry
characteristic of our model at half-filling. As usual, we include the
residue $Z$ in the coupling constant and replace the fermionic action
by the action of four components of free fermions. The effective
action describing the fermions is then given by  
\begin{widetext}
\bea
S_{\rm f} &=& a_\perp^2\int
\frac{d\tau\ d^3{\bf k}}{(2\pi)^3}
\left[R^*_{a,\alpha}(\tau,{\bf k})\left(\partial_{\tau} -
  E^R_a({\bf k})\right)R_{a,\alpha}(\tau,{\bf k})
+ L^*_{a,\alpha}(\tau,{\bf k})
\left(\partial_{\tau} - E^L_a({\bf k})\right)
L_{a,\alpha}(\tau,{\bf k})\right],
\label{fermion} 
\eea
\end{widetext}
where ${\bf k}=(q,{\bf k}_\perp)$, $\alpha=\uparrow,\downarrow$, $a=e,h$
and
\be
E_{e}^{R,L}({\bf k})=E(\pm q,{\bf k}_\perp)\ ,\quad
E_{h}^{R,L}({\bf k})=-E(\mp q,{\bf k}_\perp).
\ee
The fermion-spin vertex is described by the action 
\bea
S_{\rm int}&=&a_\perp^4\int\frac{d\tau\ d^3{\bf k}\ d^3{\bf k'}}
{(2\pi)^6}{\cal L}_{\rm int}\ ,
\eea
where 
\bea
{\cal L}_{\rm int}&=&
I_{{\bf k},{\bf k'}}\sum_{a=e,h}R^*_{a,\alpha}(\tau,{\bf k})L_{a,\beta}(\tau,{\bf k'})
g_{\alpha\beta}(\tau,{\bf k} - {\bf k'})\nn
&+&I_{{\bf k},{\bf Q}+{\bf k'}}R^*_{e,\alpha}(\tau,{\bf k})L_{h,\beta}(\tau,{\bf k'})
g_{\alpha\beta}(\tau,{\bf k} - {\bf Q}-{\bf k'})\nn
&+&I_{{\bf Q}+{\bf k},{\bf k'}}R^*_{h,\alpha}(\tau,{\bf k})L_{e,\beta}(\tau,{\bf k'})
g_{\alpha\beta}(\tau,{\bf k} + {\bf Q}-{\bf k'})  \nn
&+&{\rm h.c.}\ ,\nn
I_{{\bf k},{\bf k'}} &=&2\pi \frac{v \zh Z_2}{m^2}
\tilde{t}_{\perp}({\bf k})\tilde{t}_{\perp}({\bf k'})\ .
\label{int}
\eea
All wave vectors in the above formulae lie close to the
non-interacting Fermi surface and therefore their longitudinal
components are small in comparison to $\pi$: $|q| \ll \pi a_0$. The
entire approach is valid only when the volume inside of the Fermi
surfaces is small. One then can neglect the momentum dependence of
the exchange constant in Eq.(\ref{int}). The sign of the exchange
constant depends on the ``pocket index'' $a, b$
\bea
I_{ab} \approx \gamma m\left(
\begin{array}{cc}
1 & -1\\
-1 & 1
\end{array}
\right),
\eea
where $\gamma$ is a constant. The interaction can be cast in the form

\begin{widetext}
\bea
\label{interaction}
{\cal L}_{int} = \gamma m \Bigl[
\sum_aR^*_{a,\alpha}({\bf k})L_{a,\beta}({\bf k}')g_{\alpha\beta}({\bf k}-{\bf k}')
- \Bigl\{R^*_{e,\alpha}({\bf k})L_{h,\beta}({\bf k}') 
+R^*_{h,\alpha}({\bf k})L_{e,\beta}({\bf k}')\Bigr\}
g_{\alpha\beta}({\bf k}-{\bf k}'-{\bf Q})\Bigr] + \text{h.c.}
\eea
\end{widetext}
The value of the coupling constant $\gamma$ can be extracted from
Eqns (\ref{jsing}) and (\ref{DOS}) by noting that it is this
interaction which gives rise to the logarithmic singularity in
$J(Q)$
\bea
\tilde{J}^{\rm sing}(0,0,{\bf Q}) \approx -2\gamma^2m^2\frac{\rho(0)}{a_0}
\ln\left[\frac{\delta t}{T}\right].
\eea
Here $\rho(0)$ is the density of states per species at the Fermi
surface of coherent fermions 
\bea
\rho(0) &=& \lim_{\omega\to 0}
a_0\int \frac{d^3{\bf k}}{(2\pi)^3}\left[-\frac{1}{\pi}{\rm Im}
G(\omega,k_F+q,{\bf k}_\perp)\right]\nn
&\approx&0.539\ \frac{a_0}{(2\pi)^2}\frac{\kappa_0^2}{v}
\left[\frac{\delta t}{t_0}\right]^\frac{1}{2}.
\label{DOS} 
\eea
The result is that close to the transition we have 
\be
\gamma\propto \sqrt{\frac{t}{m}}\ ,
\ee
where $t\gg m$ is the hopping along the chains and the constant of
proportionality is of order 1. Though $\gamma$ is never small, the
small parameter $\kappa_0^2$ appears every time one integrates over
the transverse momentum. Hence the magnitude of $\gamma$ is not a
problem. 
The effective action describing the metallic side of the
Mott-insulator to metal transition is given by Eqns (\ref{spin}),
(\ref{fermion}) and (\ref{interaction}).
We find it instructive to write it down also in position space
\begin{widetext}
\bea
S_{\rm f}&=&\int d\tau d^3{\bf x}\ \Psi^\dagger_\alpha(\tau,{\bf x})
\left\{(I\otimes I)\p_{\tau} + (I\otimes\tau^z)
\left[E_0+\frac{\partial_x^2}{2M_\parallel}+\frac{\vec{\nabla}_\perp^2}{2M_\perp}
\right]\right\}\Psi_\beta(\tau,{\bf x})\ ,\nn
S_{\rm int}&=&\frac{\gamma m}{2}\int d\tau d^3{\bf x}\ \Psi^\dagger_\alpha(\tau,{\bf x})\left(
\left\{\tau^+\otimes[\exp(-2iq_0x\tau^z)-\tau^x\exp(-i{\bf Q}\cdot{\bf x}_\perp)]\right\}
g_{\alpha\beta}(\tau,{\bf x}) + {\rm h.c.}\right)\Psi_\beta(\tau,{\bf x})\ .
\eea
\end{widetext}
Here we have taken the continuum limit in the directions perpendicular
to the chains and introduced a field
$\Psi^+_\alpha = (\phi^* R^+_{e,\alpha}, \phi R^+_{h,\alpha},
\phi L^+_{e,\alpha}, \phi^* L^+_{h,\alpha})$, where
$\phi=\exp(iq_0x)$. We employ a tensor-product notation, where 
the first space is associated with the ``right/left'' index and the
second space with the ``e/h'' index. The Fermi surfaces of electrons and holes
are shifted to the origin and superimposed. The spin action $S[g]$ is
given by Eq.(\ref{spin}). Alternatively, one may use the Abelian 
representation given by Eq.(\ref{SGM}), with $g$ defined by 
(\ref{gfab}).    

\subsection{Marginal Fermi liquid}
As we shall now demonstrate, at temperatures higher than the N\'eel
temperature $T_c$ this metal is, in fact a Marginal Fermi Liquid
\cite{MFL}. The following discussion closely parallels the analysis
given by Chubukov {\it et al} for the spin-fermion model (see, for
example Ref.[\onlinecite{chub2}]). Let us consider diagrams for the
Green's function of right moving electrons. We expand around uncoupled
chains and take both the spin-fermion coupling and interchain
spin-spin exchange into account perturbatively. The elements of the
diagram technique for the fermionic degrees of freedom are as usual,
whereas the building blocks in the spin sector are the connected
2n-point spin correlators for a single chain. In diagrams that do not
contain closed fermionic loops or the interchain exchange such as the
ones in Figs 7a and 7b, the spin correlations are independent of the
transverse wave vector. This means that each fermion Green's function
is integrated over $k_{\perp}$. This integral does not differ
significantly from the integral over all momenta and as a result
is independent of $q_{\parallel}$, corresponding to a Green's function
that is local in real space: 
\be
\int \frac{\rd {\bf k}_{\perp}^2}{(2\pi)^2} \frac{1}
{\ri\omega -E^L_e(q,{\bf k}_\perp)}\approx {\rm const}\
\ri\kappa_0^2\ \mbox{sgn}(\omega)\ .
\label{local}
\ee
As (\ref{local}) is independent of $q$, we may integrate the spin
correlator in the diagram of Fig. 7a over $q$. This makes the spin
correlator local. As a result the contributions to the self energy
which do not contain closed fermionic loops or interchain spin
exchange are approximately momentum independent.  Then the self energy
calculation becomes essentially a local problem like
the problem of electron-phohon interactions in metals and
superconductors (the Eliashberg theory)\cite{sima}. In fact, such an
approach works under less stringent conditions, namely, when the spin
excitations in the transverse direction are much slower than the
quasi-particles.  Therefore the  diagrams generating a ${\bf
k}_{\perp}$ dependence of the spin-spin 
correlators, such as the ones in Fig. 7c) and
Fig. 7d) do not affect the result for the electron self energy even
close to the transition.  Once such diagrams are neglected, we get an
expansion where a factor $\kappa_0^2$ is associated with each
fermionic line (originating from the integration over ${\bf
k}_{\perp}$, as in Eq.(\ref{local}). Since $\Sigma$ depends only on
frequency, making these lines fat does not change the result
(\ref{local}) and no self-consistency is required.  
\begin{figure}[ht]
\begin{center}
\epsfxsize=0.45\textwidth
\epsfbox{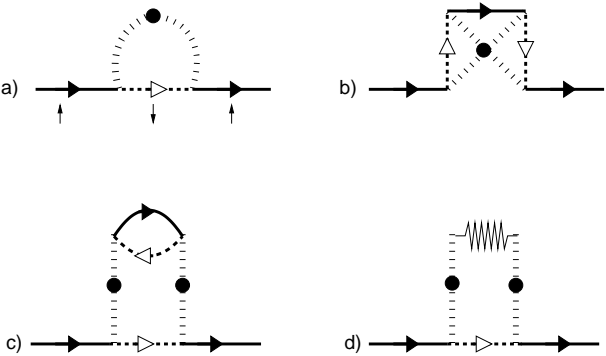}
\end{center}
\caption{Diagrams for the quasi-particle self energy of right-moving
electrons. The lines with arrows represent the fermionic Green's
functions of right and left moving electrons and holes. The 2n-point
vertices denote cumulants of the matrix fields $g$ and $g^\dagger$.}  
\label{fig:M}
\end{figure}
The contribution from the diagram Fig. 7a contains the correlation
function  
\be
\bigl\langle\!\bigl\langle
e^{\frac{i}{2}\Theta(\tau_1)}\
e^{-\frac{i}{2}\Theta(\tau_2)}
\bigr\rangle\!\bigr\rangle 
\simeq \frac{\Bigl(\ln\Bigl[\frac{\Lambda}{T}\Bigr]\Bigr)^{1/2}\pi Tv^{-1}}
{|\sin(\pi T [\tau_1-\tau_2])|}.
\ee
The contribution of the diagram in Fig. 7a) to the self-energy is then
\bea
&& \Sigma^{(a)}(\omega) \propto \kappa_0^2\Bigl(\ln\Bigl[\frac{\Lambda}{T}\Bigr]\Bigr)^\frac{1}{2} 
\int \rd \tau \frac{\re^{\ri\omega \tau}\ T}{\tau|\sin(\pi T\tau)|} \nonumber\\
&&\sim \ri\kappa_0^2
\omega\ln\left[\frac{E_0}{\mbox{max}\{\omega,T\}}\right]
\Bigl(\ln\Bigl[\frac{\Lambda}{T}\Bigr]\Bigr)^\frac{1}{2}.
\label{mfl} 
\eea
Here $E_0^{-1}$ serves as a short-time cutoff in all integrals. 
The diagram in Fig 7b) involves cumulants of the type
\bea
&&\la\!\la 
e^{\frac{i}{2}\Theta(\tau_1)}
e^{-\frac{i}{2}\Theta(\tau_2)}
e^{\frac{i}{2}\Theta(\tau_3)}
e^{-\frac{i}{2}\Theta(\tau_4)}\ra\!\ra
\eea
and gives a contribution
\bea
&&\Sigma^{(b)}(\omega)\propto \kappa_0^6
\int d\tau_2 d\tau_3 d\tau_4
\frac{e^{i\omega \tau_{14}}}{\tau_{12}\tau_{23}\tau_{34}}\nn
&\times&
\left[\left|\frac{\tau_{13}\tau_{24}}
{\tau_{12}\tau_{14}\tau_{23}\tau_{34}}\right| 
- \frac{1}{|\tau_{12}\tau_{34}|} - 
\frac{1}{|\tau_{14}\tau_{23}|}\right] \propto \omega^2
\label{4point}
\eea
to the self energy. Various other contributions are zero because some
local cumulants vanish
\bea
&&\la\!\la 
e^{\frac{i}{2}\Phi(\tau_1)}
e^{-\frac{i}{2}\Theta(\tau_2)}
e^{\frac{i}{2}\Theta(\tau_3)}
e^{-\frac{i}{2}\Phi(\tau_4)}\ra\!\ra=0.
\eea
Equation \r{4point} shows  that contributions from higher cumulants
can be neglected at small frequencies. As a result, the only essential
contribution to the self energy comes from the diagram Fig. 7a and is
given by Eq. (\ref{mfl}).

\section{ The ordered state}
As we have discussed in the previous sections, the system undergoes
an antiferromagnetic transition at a temperature much smaller than
the Mott-Hubbard gap $T_c \sim  m\kappa_0^2$. Once $T_c$ becomes small
compared to the quasi-particle Fermi energy $E_0$, one can distinguish
between metallic and insulating behavior. As we have demonstrated,
the corresponding metal is rather unusual, being in fact a Marginal
Fermi Liquid. Below $T_c$ however, the system becomes either an
insulator (for zero doping) or an ordinary Fermi liquid. Indeed, at
zero doping the electron and hole Fermi surfaces are nested and the 
ordering occurs at the antiferromagnetic wave vector in the transverse
direction (recall that the chains run along the $z$-axis) such that  
\be
\langle g_{\bl,\alpha\beta}(x)\rangle=\vec{\s}_{\alpha\beta}\cdot{\bf
  M}(-1)^{l_x+l_y}\ . 
\label{LRO}
\ee
Here the components of $\vec{\s}$ are the Pauli matrices
and ${\bf M}$ is the ordering vector.
In the mean-field approximation the fermionic spectrum is gapped
\bea
\omega^2_a = (E^R_a({\bf k}))^2 + \gamma^2 m^2 |{\bf M}|^2\ , \ a=e,h.
\eea

At non-zero doping our approach still holds provided the chemical
potential lies inside the Mott-Hubbard gap. There is no nesting any
longer and the magnetic ordering does not open a gap in the
quasi-particle spectrum. As usual, magnetic fluctuations interact
with quasi-particles through gradient vertices and these interactions
are weak.  

\section{Conclusions}
The main result of this paper is a formulation of a self-consistent 
description  
of the hybrid state of 3D quasi-particles interacting with magnetic
collective modes. The derivation is done for 
 a toy model of half filled Hubbard chains weakly coupled
through a long range interchain hopping. A certain artificiality of
the model was necessary to ensure the self-consistency of our approach
through the presence of a small parameter $\kappa_0^2$. 
We have also neglected the long-range component
of the Coulomb interaction, which plays an important role in
determining the character of the metal-insulator transition. In
reality a long-range interaction may lead to an instability of  the
small FS phase, though for small Mott-Hubbard gap its influence is
diminished by the presence of a large dielectric constant. In this
case the first order MI transition line may terminate below the
antiferromagnetic transition line of Fig. 1.  

The resulting low energy effective theory is of the Eliashberg type:
the  interaction between quasi-particles and collective modes leads to
strong retardation effects resulting in a strongly frequency
dependent quasi-particle self energy. In the present model this takes
place not just at the 'hot spots', as in  the spin-fermion model of
Chubukov and Pines, but on the entire quasi-particle FS. This makes  the
present model a candidate for the description of 'bad' metals. The
fact that in our model the electron self energy  is of the Marginal
Fermi Liquid form  is not universal and is determined by the particular
spin fluctuation spectrum.

As we discussed in the introduction, our theory provides an example
of a state where the number of carriers is unrelated to the volume
of the FS. Though this idea is well established (see e.g. the textbook
[\onlinecite{AGD}]), 
its microscopic realization was restricted to superconductors (the
example given in Ref.[\onlinecite{AGD}]). Our model provides another
example. It also demonstrates that one does not need exotic ground
states to have a small FS, as was suggested in Refs
[\onlinecite{senthil1}], [\onlinecite{senthil2}]. The 
small FS phenomenology can be generalized beyond our model. In general
there is no {\it a priori} reason for the Fermi surface even to be 
closed; for instance, Ref.[\onlinecite{rice}] describes a state with a
truncated Fermi surface observed in ARPES experiments on 
undoped cuprates \cite{arpes}.

\acknowledgments

We are grateful to Andrei Chubukov for numerous discussions and
interest to the work. AMT is grateful to S. Buehler-Paschen for
information about Hall effect measurements. AMT acknowledges support
by the US Department of Energy under contract number DE-AC02-98 CH
10886. FHLE is supported by the EPSRC under grant GR/R83712/01 and the
Institute for Strongly Correlated and Complex Systems at BNL.

\end{document}